\documentclass[%
 aip,
cp,  
 amsmath,amssymb,
 reprint,%
]{revtex4-2}

\usepackage{graphicx}
\usepackage{dcolumn}
\usepackage{multirow}
\usepackage{bm}

\usepackage[utf8]{inputenc}
\usepackage[T1]{fontenc}
\usepackage{mathptmx} 
\usepackage{comment}
\usepackage{amssymb}
\usepackage{xcolor}

\newcommand{\pd}[2]{\frac{\partial {#1}}{\partial {#2}} }

\begin{document}

\title{Radon Emanation Techniques and Measurements for LZ}

\author{N. I. Chott} 
 \email[Corresponding author: ]{nicholas.chott@sdsmt.edu}
\affiliation{South Dakota School of Mines and Technology, Rapid City, SD 57701-3901, USA}
\affiliation{For the LUX-ZEPLIN collaboration}

\author{R. W. Schnee} 
\affiliation{South Dakota School of Mines and Technology, Rapid City, SD 57701-3901, USA}
\affiliation{For the LUX-ZEPLIN collaboration}

\date{\today} 

\begin{abstract} 
Radon emanation was projected to account for  
$>50$\% of the electron recoil 
background in the WIMP region of interest for the LUX-ZEPLIN (LZ) experiment. 
To mitigate the amount of radon inside the detector volume, materials with inherently low radioactivity content were selected for LZ construction through an extensive screening campaign. The SD Mines radon emanation system was one of four emanation facilities utilized to screen materials during construction of LZ. SD Mines also employed a portable radon collection system for equipment 
too large or delicate to move to a radon emanation facility. This portable system was used 
to assay the Inner Cryostat Vessel in situ at various stages of detector construction,
resulting in the 
inference that the titanium cryostat is the source of significant radon emanation.  
Assays of a $^{228}$Th source confirmed that its $^{222}$Rn emanation is low enough for it to be used, and that 14\% of the $^{220}$Rn emanates from the source at room temperature.
\end{abstract}

\maketitle

\section{\label{section1:level1} Introduction}
Radon provides a dangerous background for 
experiments in search for WIMP dark matter~\cite{PhysRevD.101.052002,XENON:2020fbs,lux2015RnBackgrounds,deap2015Rn}. 
Radon emanation accounted for 
$>50$\% of the projected electron recoil
background in LZ~\cite{PhysRevD.101.052002,Eur.Phys.J.C80(2020)11.1044}, 
dominated by the 
“naked” $\beta$-emission from its $^{214}$Pb progeny.
For this reason nearly all components that touch the xenon were screened for radon emanation. Over eighty samples were assayed,  including 
components from the inner cryostat (such as phototubes, cabling, and PTFE), the xenon tower (such as the sub-cooler, weir reservoir, and heat exchanger), and the xenon circulation system (such as compressors, circulation panel, and xenon transfer lines)~\cite{Eur.Phys.J.C80(2020)11.1044}.  
Here we briefly summarize the process, detail the particularly interesting and complicated emanation measurement of the full Inner Cryostat Vessel (ICV), and describe radon emanation of a calibration source not included in Ref.~\cite{Eur.Phys.J.C80(2020)11.1044}.

\section{\label{section:Methods} Radon Emanation Measurements}

\begin{table} 
\caption{Comparison of the 
four radon emanation facilities used by LZ. The chambers listed 
contain the sample material, where radon is collected. 
The emanation rates from the chambers alone (the "blank" rates) are statistically subtracted for sample measurements. Only a fraction of the emanated radon is transferred to the detection chamber ("Transfer Eff.") and only a fraction of that transferred is then detected ("Detector Eff."). 
The cross-calibration figures represent the reconstructed emanation rate of a standard rubber sample previously used by 
the EXO collaboration. When not stated, overall uncertainties are estimated to be 10--20\% (consistent with most of the cross-calibrations).}
\label{tab:facilities}
    \tabcolsep=6pt
    \centering
    \begin{ruledtabular}
    \begin{tabular}{ ccccccc }
     \multicolumn{1}{c}{Detector} & 
    \multirow{2}{*}{Type} &  
    Chamber &   
    Chamber Blank &  
    Transfer & 
    Detector & 
    Cross-Calibration \\ 
    & 
     & 
    Volumes [L] & 
    Rates [mBq] & 
    Eff. [\%] & 
    Eff. [\%] & 
    [Measured/EXO activity]\\ 
    \hline
    \vspace{-4mm}
    \\ 
    SD Mines & 
    PIN-diode & 
    \multicolumn{1}{c}{\begin{tabular}[c]{@{}c@{}}13\\ 300\end{tabular}} & \multicolumn{1}{c}{\begin{tabular}[c]{@{}c@{}}0.2\\ 0.2\end{tabular}} & 
    \multicolumn{1}{c}{\begin{tabular}[c]{@{}c@{}}94\\ 80\end{tabular}} & 
    25 & 
    \multicolumn{1}{c}{\begin{tabular}[c]{@{}c@{}}0.89 $\pm$ 0.15\\ 1.11 $\pm$ 0.28\end{tabular}}
    \vspace{2mm}
    \\ 
    Maryland & 
    PIN-diode & 
    4.7 & 
    0.2 & 
    96 & 
    24 & 
    1.13 $\pm$ 0.19
    \vspace{2mm} \\
    UCL & 
    PIN-diode & 
    \multicolumn{1}{c}{\begin{tabular}[c]{@{}c@{}}2.6\\ 2.6\end{tabular}} &
    \multicolumn{1}{c}{\begin{tabular}[c]{@{}c@{}}0.2\\ 0.4\end{tabular}} & 
    \multicolumn{1}{c}{\begin{tabular}[c]{@{}c@{}}97\\ 97\end{tabular}} & 
    30 & 
    1.49 $\pm$ 0.15
    \vspace{2mm} \\ 
    Alabama & 
    Liquid Scint. & 
    \multicolumn{1}{c}{\begin{tabular}[c]{@{}c@{}}2.6\\ 2.6\end{tabular}} &
    \textless{}0.4 & 
    34 &
    36 & 
    0.83 $\pm$ 0.17
    \\
\end{tabular}
\end{ruledtabular}
\end{table}

Four facilities, detailed in Table~\ref{tab:facilities}, performed the LZ radon emanation measurements.  For each, emanation of the materials took place in dedicated chambers formed from electropolished stainless steel in order to minimize emanation from the chambers themselves, with the ``blank'' rate from the chamber alone typically 0.2--0.4\,mBq.  Leak checks ensured that radon from lab air would not enter the chamber during the emanation period and provide additional background.
All facilities operated at room temperature such that the expected suppression of diffusion-dominated radon emanation at low temperature was not probed.
 
While transfer from small emanation chambers to the detection chamber was relatively straightforward, high transfer efficiency was obtained even from the large SD Mines emanation chambers.
A low-radon carrier gas, such as nitrogen or helium, was passed through a cold trap consisting of activated carbon cooled with a solution of dry ice and isopropyl alcohol to render its radon concentration negligible.  The carrier gas passed into the emanation chamber, then through a second cold trap.
The second cold trap was made of brass or copper wool cooled to 77K by submerging it in liquid nitrogen and was used to trap the radon atoms exiting the emanation chamber while allowing the carrier gas to be pumped away. 
In order to trap a high percentage of the radon atoms from a large emanation chamber, the chamber was pumped down with its gas passing through the metal trap, then refilled with filtered carrier gas, with the process repeated five times.

The Alabama facility collects the harvested radon by dissolving it in organic liquid scintillator by means of a carrier gas. 
The delayed $^{214}$Bi-$^{214}$Po coincidences are then counted to infer the $^{222}$Rn decay rate.  The other three facilities all use electrostatic PIN-diode detectors, where the detector is at a negative voltage relative to the detection chamber.  Since most of radon daughter nuclei are positively charged  ($87.3 \pm 1.6$\% in air)~\cite{PAGELKOPF20031057}, about half of the daughters end up on the PIN-diode itself. Half of the resulting alpha decays deposit the alpha energy in the detector, resulting in detector efficiencies about 25\% as listed in Table~\ref{tab:facilities}.  Because $^{218}$Po$^+$, $^{214}$Pb$^+$, and $^{214}$Bi$^+$ ions may all be collected on the detector, $^{214}$Po has a slightly higher collection efficiency than $^{218}$Po, with calibrations at SD Mines indicating a detection efficiency of 23\% for $^{218}$Po and 26\% for $^{214}$Po for its detector~\cite{MillerLZbgdsLRT2018}.

\begin{figure}[tb]
    \centering
    \includegraphics[width=0.97\textwidth]{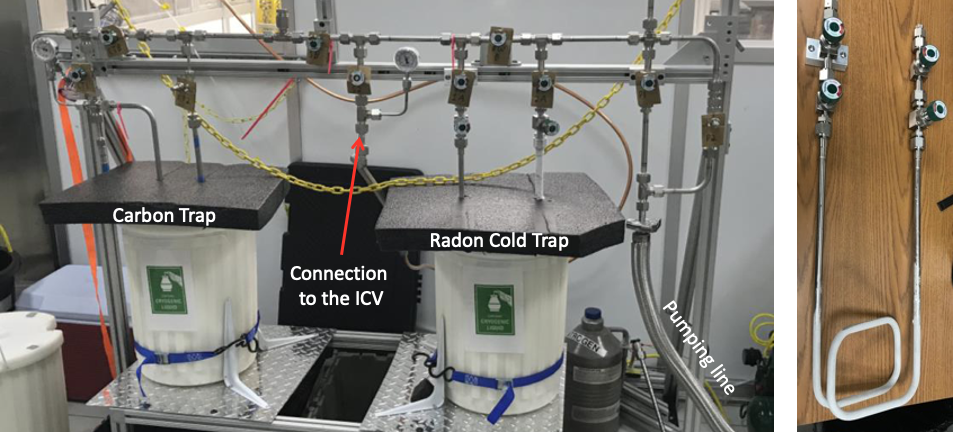}
    \caption{\label{fig:PortableTrap} 
    \textbf{Left:} SD Mines portable radon system at 
    the Sanford Underground Research Facility (SURF) Surface Assembly Lab for an early measurement of the Inner Cryostat Vessel (ICV), before the full inner detector was completed.  Liquid nitrogen (LN) boil-off gas is cleaned of residual radon in the cooled Carbon Trap before flowing into the ICV and then out to the Radon Cold trap for harvesting.
    \textbf{Right:} Portable Radon Cold Trap itself disconnected for transport to the SD Mines campus.
   }
\end{figure}

LZ used two portable radon collection systems (Maryland and SD Mines), for equipment too large or delicate to move to the radon emanation facilities, or for equipment used as their own emanation volumes. 
Figure~\ref{fig:PortableTrap} shows the SD Mines portable system at the Sanford Underground Research Facility (SURF).
After the transfer of radon to the portable cold trap, it was double-sealed with hand valves, warmed, and removed from the portable system. The trap was then transported to a radon screening facility where the collected radon was transferred  from the trap into a detection chamber for counting.  The Maryland portable system was used to determine that the xenon circulation system's integrated compressor skid assembly originally presented $\sim$17\,mBq. Replacing most of the welded stainless steel plumbing and etching the accumulation bottles in citric acid reduced the rate to $1.48\pm0.31$\,mBq~\cite{Eur.Phys.J.C80(2020)11.1044}.  The SD Mines portable system was used to assay the fully loaded getter (model PS5-MGT50-R-535 from SAES) at its operational temperature of $400^{\circ}$C using helium carrier gas; its emanation rate was determined to be $2.26^{+0.28}_{-0.27}$\,mBq~\cite{Eur.Phys.J.C80(2020)11.1044,CarterHallLRT2022}.  The SD Mines portable system was also used to assay other large equipment at SURF such as the Xenon Tower and the Inner Cryostat Vessel.
\section{\label{sec:ICV}
Inner Cryostat Vessel with Full Detector}
\begin{figure}[tb]
    \centering
    \includegraphics[width=0.8\textwidth]{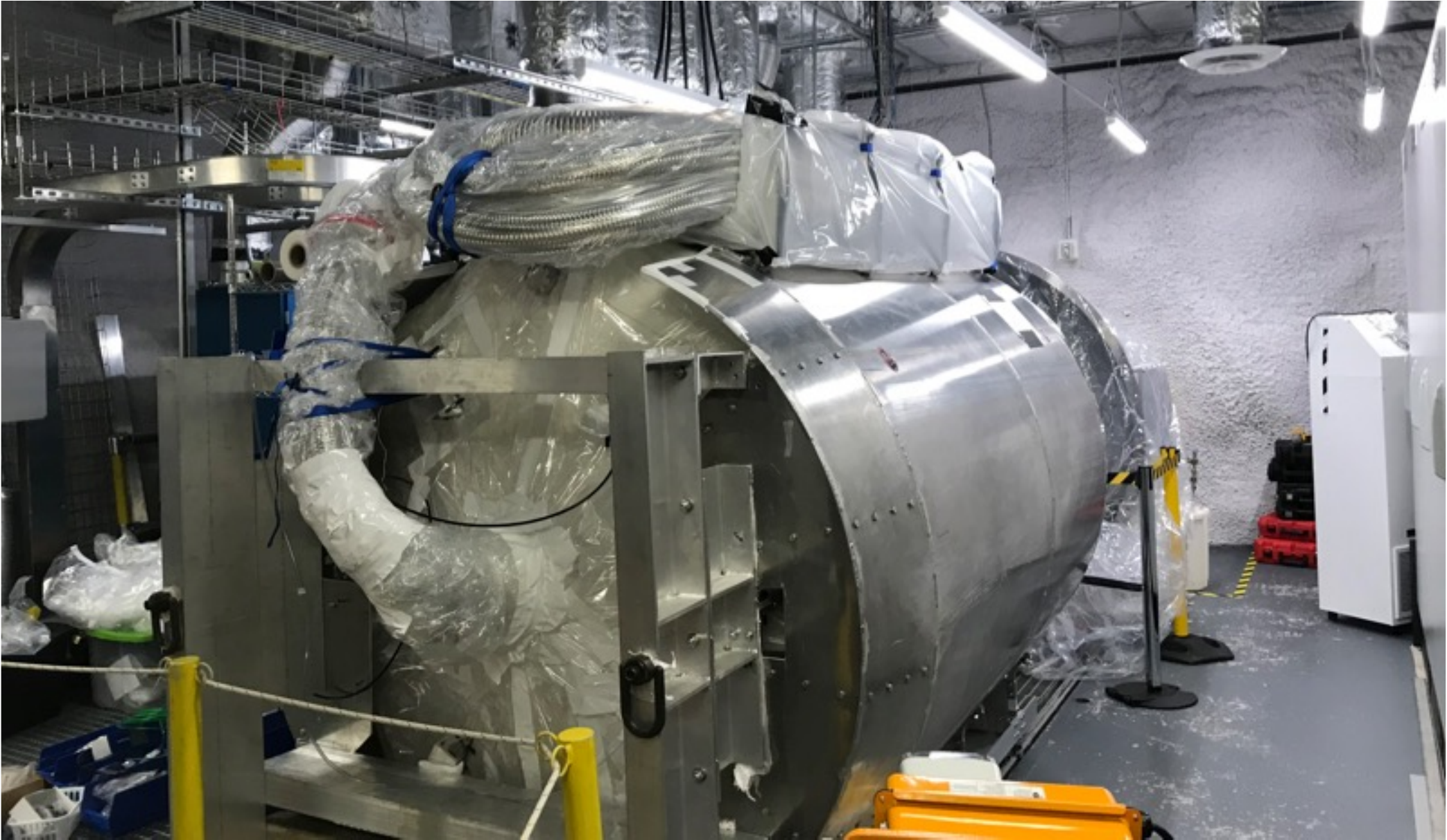}
    \caption{\label{fig:ICVphoto} Photo of ICV in October 2019 after it was enclosed with all components of the inner detector, filled with low-radon nitrogen gas, and moved underground to the 4850-foot level of the Sanford Underground Research Facility,  allowing it two weeks to emanate.  Once underground, the SD Mines portable emanation system was used to collect a fraction of the emanated radon.  
   }
\end{figure}

Measuring  the radon emanation from
the Inner Cryostat Vessel (ICV) after the assembly of the full inner detector
was complicated.
Radon emanation from the ICV had been measured several times during the integration of various detector components. The final assay was made in October 2019 after the ICV was fully complete and sealed as shown in Fig.~\ref{fig:ICVphoto}. The cryostat at this stage housed the entire inner detector including photomultiplier tube arrays, their corresponding bases and cables, the entire field cage, PTFE coating, sensors, and conduit volumes.
The South Dakota Mines portable radon trapping system was deployed underground at SURF with minimal plumbing due to space constraints. After leak-checking and purging, the trapping system was opened to the ICV and the emanated gas was harvested over a 6.3-hour period that removed 18.3\% of the gas within the ICV. After the harvest, the trap was carefully disconnected and transported to the SD Mines radon facility for screening. 
\begin{figure}[tb]
    \centering
    \includegraphics[width=\textwidth]{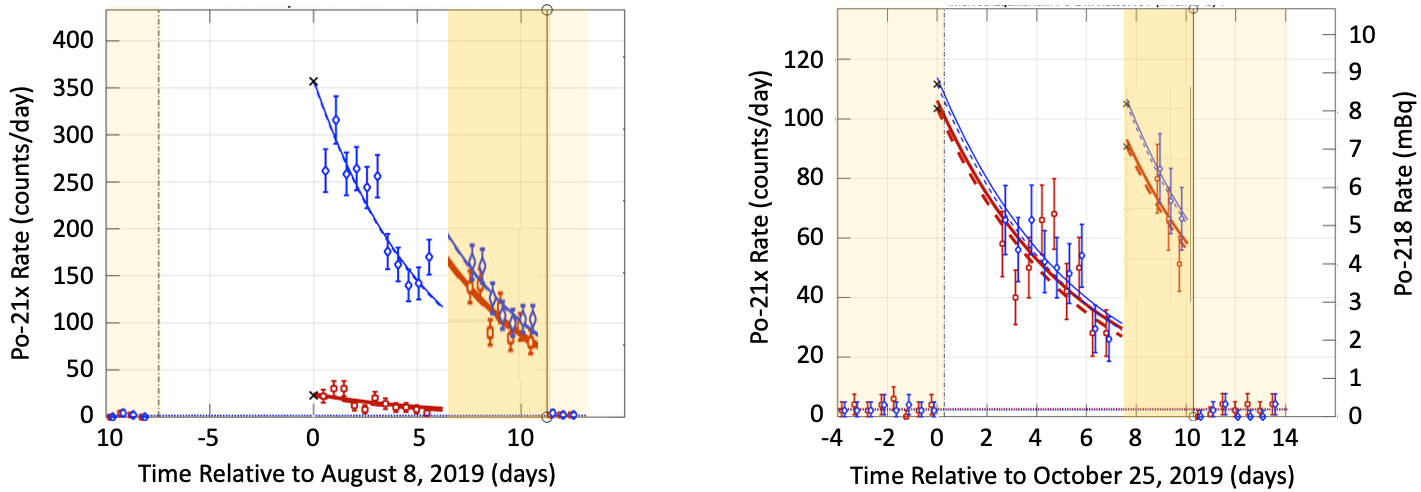}
    \caption{\label{fig:ICVresults} Detected rates for $^{218}$Po (red diamonds) and $^{214}$Po (blue circles) as functions of time relative to sample transfer, with best-fit  $^{222}$Rn curves decay curves for sample (unshaded) and calibration runs (darkly shaded), and best-fit constant values for background runs (lightly shaded).  
    \textbf{Left:} Test measurement with $87\pm4$\,mBq added to the detection chamber after an early transfer of radon and contaminants from the IVC performed without the filtering process.  The low rates detected relative to those expected, and the lower rate of $^{218}$Po relative to $^{214}$Po, both indicate neutralization of positive ions before they reached the detector surface. 
    After the filtration procedure  and a step believed to lose 2/3 of the radon sample, the detection rate of $^{214}$Po increased by 50\% and the detection rate of $^{218}$Po increased by $22\times$ (darkly shaded).
    \textbf{Right:} Measurement of ICV with completed inner detector after filtering process.  Agreement of $^{218}$Po and $^{214}$Po rates suggests a normal collection efficiency for each.  The measurement of additional radon added to the detection chamber (darkly shaded) is also consistent with a normal collection efficiency. 
   }
\end{figure}

Earlier emanation measurements of the ICV had indicated that the radon trap also captured an outgassed molecular species that neutralized the positively charged radon daughters in the radon detection chamber, leading to a drop in detection efficiency.
A residual gas analyzer (RGA) indicated the culprit had 59 AMU, but no clear candidate was identified.
The left panel of Fig.~\ref{fig:ICVresults} shows a test measurement made in August 2019 by adding a known amount of radon from a Pylon radon source to the gas transferred from an earlier assay of the ICV.  For this test, a sample of $87\pm4$\,mBq was transferred to a cold trap and the trap was warmed.  The gas from the trap was transferred into the detector volume and allowed to equilibrate within the full volume containing the trap, tubing, and detector.  Since the detector chamber volume is 97\% of this total (and based on previous calibrations),  $>95$\% of the radon is expected to have transferred into the chamber.  However, the detected rate for $^{214}$Po is about $4\times$ lower than that expected for this amount of radon, while the detected rate of $^{218}$Po is $>50\times$ lower than expected.  These results indicate that $^{218}$Po$^+$ is neutralized before reaching the detector surface $>98$\% of the time.  The much higher detected rate of $^{214}$Po decays presumably results from the ions $^{214}$Pb$^+$ and $^{214}$Bi$^+$ having a lower neutralization probability than $^{218}$Po$^+$.

To increase the collection efficiency, a procedure was developed to remove the contaminants from the gas while keeping the radon by taking advantage of the fact that radon breaks through our brass-wool cold trap very quickly when it is held at $-78^{\circ{}}$C (with a mixture of dry ice and IPA), but contaminants may break through the trap more slowly.  The sample 
was transferred from a brass-wool cold trap held at $-78^{\circ{}}$C to one at $-196^{\circ{}}$C (LN) with sufficient flow to transfer all of the radon atoms while leaving most of the contaminants behind, as confirmed with the RGA.  This cold trap was then purged. 
The radon sample was transferred to the detection chamber via a secondary small cold trap.  
As shown in the left panel of Fig.~\ref{fig:ICVresults},
detection of a consistent, increased rate of both $^{214}$Po and $^{218}$Po  strongly suggests that the process was successful at removing the contaminants.  Measurements with the RGA also indicated a significant reduction in contaminant concentration at 59 AMU.
Unfortunately, the transfer of contaminants to the RGA involved flowing LN boiloff gas through the trapped radon, 
losing $\sim$67\% of the radon sample.   

This process was used, without the step that likely lost sample, for the transfer of the sample from the completed ICV.
The right panel of Fig.~\ref{fig:ICVresults} shows the results.  The agreement of the $^{214}$Po and $^{218}$Po rates suggest that the process was successful and detection efficiencies were therefore the usual 23\% for $^{218}$Po and 26\% for $^{214}$Po.   
Immediately following the measurement of the ICV radon sample, a check on the detection efficiency was attempted by adding radon from the Pylon source to the gas already in the detection chamber.  Unfortunately, only a small amount of radon, with a large systematic uncertainty, was added.  Results taken at face value indicate a higher detection efficiency than usual.  However, since there is no plausible way to have increased the detection efficiency above the standard values, it is more likely that the systematic uncertainty on the radon added was underestimated.  
We therefore use these usual detection efficiencies for calculation of the ICV results.

Based on the observed $^{214}$Po and $^{218}$Po rates, the radon activity of the sample  was  $8.07^{+0.62}_{-0.59}\,$mBq.  Transfer efficiencies were dominated by the small fraction of gas removed from the ICV, with an additional 0--8\% of the collected sample lost to post-filter RGA measurements.  Under the assumption of an even sampling of the radon within the ICV, the total transfer efficiency was $17.5\% \pm 0.7$\%, resulting in a total radon emanation rate of  $46.1^{+4.0}_{-3.8}\,$mBq for the ICV including all materials within it.  

The expected radon contribution from all materials inside the ICV, excluding dust and the ICV titanium itself, was $27.2 \pm 1.9\,$mBq, while dust could plausibly contribute anywhere from 1.7 -- 9.1\,mBq.  The ICV titanium therefore must contribute 10--20 mBq of radon burden.  Since the total titanium surface area in the ICV is 15.1\,m$^2$, the inferred emanation is $\sim$1\,mBq/m$^2$, consistent with measurements of LZ titanium sheets at UCL~\cite{UmitThesis}. 
Assays with high-purity germanium detectors had set an upper limit of $<0.09$\,mBq/kg on $^{226}$Ra in the  800-kg ICV~\cite{Eur.Phys.J.C80(2020)11.1044}, so the maximum radon burden for it (assuming half emanates to the outside and half to the inside) is 36\,mBq. These measurements suggest a high fraction of the radium is very near the titanium surface (or that radon diffuses surprisingly well through titanium).  Previous measurements of titanium by XENON1T~\cite{XENON:2020fbs} included one set of samples (total area 1.1\,m$^{2}$ and mass 4.6\,kg) with 5.0\,mBq $^{226}$Ra and 3.0\,mBq $^{222}$Rn emanation, so such a high fraction of emanation in titanium has been seen before.  
For these samples, electropolishing 30\,$\mu$m off the surface reduced the emanation by $>300\times$ to a result consistent with zero, indicating that the  $^{226}$Ra had indeed been predominantly on the surface.
The assay results for the LZ ICV similarly suggest that most of the $^{226}$Ra is near the titanium surfaces.  
Since the radon emanation from titanium is of great interest to future experiments, further studies are warranted.

\section{\label{sec:Th228source}%
LZ $^{228}$Th Calibration Source}

A  $^{228}$Th calibration source ($\tau_{1/2}=1.9116$\,y), used to inject short-lived $^{220}$Rn into the LZ TPC for calibration, was measured to make sure its $^{222}$Rn emanation was not too large.  The source was nominally 20.11\,kBq on May 1, 2020, and is enclosed in stainless steel tubing between two 3-nm pore-sized filters between two locking valves.  Emanation was performed twice successfully from December 2020 to January 2021 using the source as its own chamber, with the valves closed for two weeks.  Radon was transferred first into the SD Mines 13-liter emanation chamber at low pressure by flowing 4\,liters of LN boiloff gas (many times the source volume) through the source.  Then the standard  procedure was used to transfer the radon to the cold trap.  The trap was isolated for 2--3 days to allow both the $^{220}$Rn and most of its longer-lived daughters to decay before transfer to the detection chamber, in case the transfer efficiency for these longer-lived daughters was high enough to otherwise overwhelm the detector.
Results indicate the $^{222}$Rn emanation  rate is $0.88 \pm 0.18$\,mBq, acceptably small compared to the LZ $^{222}$Rn background to allow its use as a calibration source.

\begin{figure}[tb]
    \centering
    \includegraphics[width=0.9\textwidth]{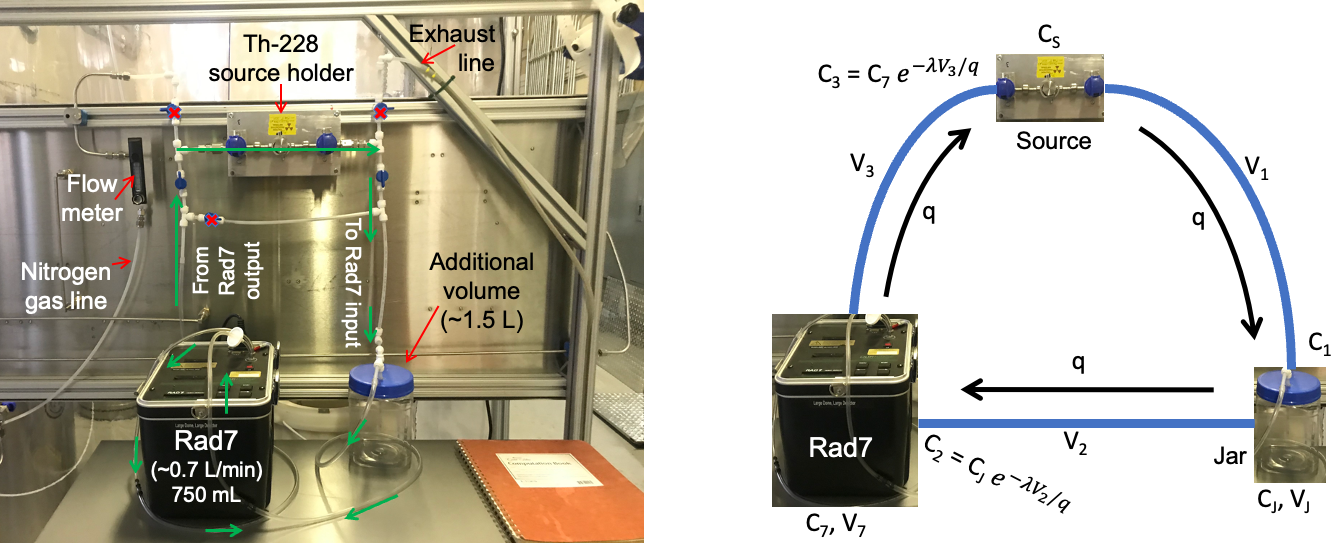}
    \caption{\label{fig:ThoronResults} \textbf{Left:} Setup for $^{220}$Rn assay of $^{228}$Th calibration source, which lives in the stainless-steel tubing just under the radioactivity warning sign, surrounded by locking valves.  LN boil-off gas flowed through the nitrogen gas line, flowmeter, source, and exhaust line to purge the source before use.  During the measurement, the RAD7's internal pump continuously circulated about 0.7 liters/minute via the path shown with arrows.  A jar provided additional volume to reduce any risk of the RAD7 pump damaging the source.  The push-to-connect unions and valves were sealed with silicone.
 \textbf{Right:} Schematic of setup labeling radon concentration variables $C_1, C_2, C_3, C_7, C_{\rm S}, C_{\rm J}$ and volumes $V_1 = 18.0$\,ml, $V_2 = 22.0$\,ml, $V_3 = 26.3$\,ml, $V_{\rm J} = 1479$\,ml, and $V_7 = 750$\,ml, and showing flow $q=0.7$\,liters/min.  The volume of the source is negligible.
   }
\end{figure}

The amount of $^{220}$Rn emanating from the source was measured using a Durridge RAD7 detector~\cite{durridgeRAD7} on February~11, 2021, when the $^{228}$Th activity would have decreased to 15.14\,kBq.  After purging the $^{228}$Th source with LN boiloff, valves were simultaneously changed to start circulating LN boiloff through the source, to a jar to provide a large (1.5\,liter) volume, and a Durridge RAD7 detector~\cite{durridgeRAD7}, as shown in 
Fig.~\ref{fig:ThoronResults}.  
In the limit of perfect mixing, the concentration $C$ in one of the large volumes (the jar or the RAD7) depends on its volume $V$, the flow $q$, the $^{220}$Rn decay rate $\lambda$, and the concentration $C_i$ 
feeding into it:
\begin{equation*}
\pd{C}{t} = \frac{q}{V} C_i - \frac{q}{V} C - \lambda{} C,
\end{equation*}
with steady-state solution
$C = q C_i / \left(q+ \lambda{}V\right)$.  
The time for radon to move from the start to the end of one of the small tubes of volume $V_i$ is $V_i/q$, so in equilibrium the radon concentration in such a tube decreases along its length from $C$ at the start of the tube to $C_i = C e^{-V_i \lambda/q}$ at the end of the tube.
The thorium source increases the concentration from its input $C_3$ to $C_3+ C_{\rm Th}$ at its output, where $C_{\rm Th} = E / q$ is the concentration in the source 
due to its emanation $E$.
Combining these equations for the setup sketched in Fig.~\ref{fig:ThoronResults} yields the expected ratio of thoron concentration in the RAD7 to that in the source:
\begin{equation*}
\frac{C_7}{C_{\rm Th}} = \frac{e^{-\left[ \left( V_1 +V_2 \right) \lambda \right] /q}} {\left( 1 + \lambda V_J/q \right) \left( 1 + \lambda V_7 / q \right) - e^{ \left[ \left( V_1 + V_2 + V_3 \right) \lambda \right] /q } } = 0.24.
\end{equation*}
The measured thoron concentration in the RAD7  $C_7 = 579.6 \pm 2.6$\,kBq/m$^3$.  Including systematic uncertainties on the flow and volume results in an inferred $^{220}$Rn emanation rate $E = 2.11 \pm 0.19$\,kBq, so about 14\% of the $^{220}$Rn escapes the source at room temperature.

The lack of $^{212}$Po alpha decays after the transfer allows limits to be placed on the fraction of $^{212}$Pb atoms that were transferred into the detection chamber. 
The strongest limit may be set from the second emanation transfer, for which the delay in the cold trap was shorter, 52\,hours.  
The two weeks of emanation put the $^{220}$Rn decay chain in excellent secular equilibrium.  At the start of the transfer, there were $E/\lambda \approx 167,000$ $^{220}$Rn atoms in the source, along with 450 $^{216}$Po atoms, $115 \times 10^6$\,$^{212}$Pb atoms, and $11 \times 10^6$\,$^{212}$Bi atoms.  At the end of the transfer, there were $N_{\rm end}=3.8 \times 10^6$\,$^{212}$Pb atoms.  Of these, $\epsilon_{\rm decay} \approx 75$\%  decayed through $^{212}$Po to $^{208}$Pb during the first day of measurement, with about $\epsilon_{\rm det}\approx25$\% of such atoms in the detection chamber resulting in the daughter $^{212}$Bi atoms collected on the detector and detected.  Since no events consistent with $^{212}$Po (or $^{212}$Bi--$^{212}$Po coincidences) were observed during the first day of measurement, the 90\% C.L. upper limit of $N_{\rm obs} \leq 2.3$, and the the fraction of the $^{212}$Pb atoms transferred to the chamber 
\begin{equation*}
\frac{N_{\rm obs}}{\epsilon_{\rm decay} \epsilon_{\rm det} N_{\rm end}} \leq \frac{2.3}{(0.75)(0.25)(3.8\times 10^6)} =  2.4 \times 10^{-6}.
\end{equation*}
Clearly, the two-day delay before transferring the sample from the trap to the detection chamber was unnecessary.

\section{Conclusions}

Radon's importance as a background for LZ required a comprehensive and sensitive radon emanation assay program.  Measurement of LZ's inner cryostat vessel indicates a large fraction of $^{226}$Ra in the titanium results in $^{222}$Rn emanation, likely due to surface contamination.  Comprehensive assay results of materials used in LZ construction  are available in Ref.~\cite{Eur.Phys.J.C80(2020)11.1044}.  One item not included in that publication, the LZ $^{228}$Th source, was found to emanate sufficiently little $^{222}$Rn to allow its use for LZ calibrations.

\begin{acknowledgments}
The authors gratefully acknowledge the support and participation of the LZ collaboration.  Ann Harrison performed or helped perform  many radon emanation measurements of the ICV and developed the filtering procedure used.  Carter Hall's group at the University of Maryland designed the portable radon emanation systems.
Scott Hertel's group at the University of Massachusetts provided the $^{228}$Th source.
This work was supported in part by the Department of Energy (Grant No.~DE-AC02-05CH1123).

\end{acknowledgments}

\nocite{*}
\bibliography{aipsamp}

\end{document}